\documentclass[]{pasj01}

\Received{}
\Accepted{}
 
\usepackage{lineno}
\begin{document} 

\title{ 
Twisted magnetic field in star formation processes of L1521\,F revealed by submillimeter dual band polarimetry using James Clerk Maxwell Telescope}

\author{Sakiko \textsc{Fukaya} \altaffilmark{1}}
\altaffiltext{1}{Physics and Astronomy Department, Graduate School of Science and Engineering,
  Kagoshima University, Kagoshima 890-0065, Japan}
\email{k7208474@kadai.jp}

\author{Hiroko \textsc{Shinnaga}\altaffilmark{1,2}}
\email{shinnaga@sci.kagoshima-u.ac.jp}
\altaffiltext{2}{Amanogawa Galaxy Astronomy Research Center(AGARC), Graduate School of Science and Engineering,
  Kagoshima University, Kagoshima 890-0065, Japan}

\author{Ray S. \textsc{Furuya}\altaffilmark{3}}
\altaffiltext{3}{Institute of Liberal Arts and Sciences, Tokushima University, Tokushima 770-8502, Japan}
\email{rsf@tokushima-u.ac.jp}

\author{Kohji \textsc{Tomisaka}\altaffilmark{4}}
\altaffiltext{4}{National Astronomical Observatory of Japan, Tokyo 181-8588, Japan}
\email{tomisaka@th.nao.ac.jp}

\author{Masahiro N. \textsc{Machida}\altaffilmark{5}}
\altaffiltext{5}{Department of Earth and Planetary Sciences, Faculty of Sciences, Kyushu University, Fukuoka 819-0395, Japan}
\email{machida.masahiro.018@m.kyushu-u.ac.jp}

\author{Naoto \textsc{Harada}\altaffilmark{5}}
\email{harada.naoto.450@s.kyushu-u.ac.jp}


\maketitle

\begin{abstract}
Understanding the initial conditions of star formation requires both observational studies and theoretical works taking into account the magnetic field, which plays an important role in star formation processes. Herein, we study the young nearby dense cloud core L1521\,F ($n$(H$_2$) $\sim 10^{4-6}$ cm$^{-3}$) in the Taurus Molecular Cloud. This dense core hosts a 0.2 $M_\odot$ protostar, categorized as a Very Low Luminosity Objects with complex velocity structures, particularly in the vicinity of the protostar. To trace the  magnetic field within the dense core, we conducted high sensitivity submillimeter polarimetry of the dust continuum at $\lambda$= 850 $\micron$ and 450 $\micron$ using the POL-2 polarimeter situated in front of the SCUBA-2 submillimeter bolometer camera on James Clerk Maxwell Tetescope.  
This was compared with millimeter polarimetry taken at $\lambda$= 3.3 mm with ALMA. The magnetic field was detected at $\lambda$= 850 $\micron$ in the peripheral region, which is threaded in a north-south direction, while the central region traced at $\lambda$= 450 $\micron$ shows a magnetic field with an east-west direction, i.e., orthogonal to that of the peripheral region. Magnetic field strengths are estimated to be $\sim$70 $\mu$G and 200 $\mu$G in the peripheral- and central-regions, respectively, using the  Davis-Chandrasekhar-Fermi method. The resulting mass-to-flux ratio of 3 times larger than that of magnetically critical state for both regions indicates that L1521\,F is magnetically supercritical, i.e., gravitational forces dominate over magnetic turbulence forces. Combining  observational data with MHD simulations, detailed parameters of the morphological properties of this puzzling object are derived for the first time.   

\end{abstract}
\clearpage

\section{Introduction}
~~~Stars are formed in the high-density cores within the interstellar molecular clouds, which consist of cold ($T= 10$ K) molecular gas and dust originating from winds and explosions of evolved low- and high-mass stars.   
The magnetic field as well as thermal and turbulence pressures interact to oppose gravitational collapse of the core. Accordingly, studying magnetic fields is crucial to understand detailed processes of star formation. \par
We here study the dense core of L1521\,F (also called MC 27) in the Taurus Molecular Cloud, which is one of the nearest sites of low mass star formation at a distance of 140 pc (\cite{Elias97}). 
The density of the central 1000 AU region of L1521\,F is estimated to be $\sim$ $10^6$ cm$^{-3}$ (\cite{Onishietal99}).
The dense core hosts L1521\,F-IRS, a protostar categorized as 
Very Low Luminosity Object (\citet{diFrancesco07}) whose luminosity is 0.05 $\LO$ according to Spitzer measurements at 3.5, 4.5, and 8.0 $\micron$ (e.g., \cite{Bourkeetal06}).
The Spitzer space telescope has detected a bipolar nebula that traces a bipolar outflow cavity stretched over 4000 AU within the core, almost on the plane of the sky. \par
CCS ($J_N =3_2 - 2_1$) and N$_2$H$^+$ ($J =1 - 0$) molecular emissions taken using 
the Berkeley-Illinois-Maryland Association Array 
indicate eight and four substructural clumps within the core, respectively.  CCS emissions ($n_{\rm crit~CCS} \sim 10^{4}$ cm$^{-3}$) trace peripheral region of the core, whereas N$_2$H$^+$ emissions ($n_{\rm crit~N_2H^+} \sim 10^{5}$ cm$^{-3}$) trace its central dense region. Velocity maps of these two molecular emissions 
indicate that there is a notable gap in velocity structures between the peripheral and central regions of the dense core (\cite{Shinnagaetal04}).  
The Caltech 10.4 m Leighton telescope of the Caltech Submillimeter Observatory 
detects high $J$ transitions of $^{12}$CO ($J= 6 - 5$ and $J= 7 - 6$) toward the center of the dense core. CO $J= 6 - 5$ 
traces Warm ($\sim 30 - 70$ K), Extended (radius of $\sim$ 2400 AU), Dense ($\sim$ a few 10$^5$ cm$^{-3}$) Gas (called WEDG),  
in the Warm-in-Cold Core Stage (WICCS; \cite{Shinnagaetal09}), which would inaugurate a missing link in evolution between a cold quiescent starless core phase and a young protostar in class 0 stage. \par
High-resolution observations acquired using Atacama Large Millimeter/submillimeter Array (hereafter ALMA) at $\lambda$= 870 $\micron$ reveal high-density condensations named MMS-1, 2, and 3 at the center of the core (\cite{Tokudaetal16}). MMS-1 corresponds to a 0.2 $M_\odot$ protostar with a 10 AU-scale disk and this is the only protostar in the system (\cite{Tokudaetal17}). The other two millimeter sources, MMS-2 and MMS-3, are minor condensations compared to MMS-1, considering their masses ranging 10$^{-3}$ to 10$^{-2} M_\odot$ (Tokuda et al. 2016). 
$^{12}$CO($J= 3-2$) shows very complex and warm (\textless ~60 K) filamentary structures around the protostar, likely generated by turbulence (\cite{Tokudaetal18}). 
\par
The global magnetic field surrounding L1521F core 
is in a north$-$south direction as measured by Planck $\nu$= 353 GHz ($\lambda$= 850 $\mu$m) observations  (\cite{Planckcollaboration15}).
Both Optical R-band polarimetry and 
850 $\micron$ dust continuum polarimetry show north$-$northeast- to south$-$southwest 
magnetic field direction in the outskirt of the dense core (\cite{Soametal15}, 2019). \par 
This paper reports the magnetic field morphology within the core measured at $\lambda$= 850 $\micron$ and 
$\lambda$= 450 $\micron$ for the first time toward this object, together with  
$\lambda$= 3.3 mm polarimetry results using ALMA. 
The SED of the core is also presented to infer parameters of the protostar and the core itself based on the method presented by Robitaille (2017). 
Comparing these observational results with MHD theoretical 
simulations described by Machida et al. (2020), we discuss the gravitational infall scenario of the core based on the course of star formation. 


\section{Observation and data reduction}
\subsection{Polarimetry with Submillimeter Common Users Bolometer Array-2 (SCUBA-2) / Polarimeter-2 (POL-2) at James Clerk Maxwell Telescope (JCMT)}
 ~~~ Submillimeter polarimetry data toward L1521\,F were taken at $\lambda$= 850 $\micron$ and $\lambda$= 450 $\micron$ with Submillimeter Common Users Bolometer Array 2 (SCUBA-2) and polarimeter (POL-2) (\cite{Hollandetal13}) mounted on the 15-m James Clerk Maxwell Telescope (JCMT). Dual band polarimetry 
 enables magnetic field morphology to be traced in the peripheral cold region of the core at $\lambda$= 850 $\micron$ as well as that in its central warm region at $\lambda$= 450 $\micron$. Daisy scans, which are suitable for  observing compact sources, were utilized to take the data. Spatial resolutions at $\lambda$= 850 $\micron$ and 450 $\micron$ were 14.1$\arcsec$ (corresponds to 1900 AU at 140 pc) and 8$\arcsec$ ($\sim$ 1080 AU), respectively (\cite{Dempseyetal12}). \par
 
 Instrumental polarization was calibrated by using the model released on August 3, 2019. 
  The instrumental polarization model includes a new de-biased method 
in order to treat the positive noise in $PI$ (polarized intensity) using a modified asymptotic estimator (Plaszczynski et al. 2014; Montier et al. 2015), which has not been applied in the past for the instrumental polarization corrections of the instrument.  The method enables us to improve significantly in the resulting S/N in $PI$ of the polarization data sets. \par
 Stokes parameters $I$, $Q$, and $U$ were used to derive the polarization fraction ($P$) and polarization angle ($ANG$),  
 which can be described as follows: 

 \begin{equation}
P=\frac{\sqrt{Q^2+U^2}}{I} \label{eq1},
\end{equation}

 \begin{equation}
ANG =\frac{1}{2} \arctan(\frac{U}{Q}) \label{eq2}.
\end{equation}
~~~ Polarization data were acquired on October 19, November 11, 17, 18, and 21, 2017.  Total integration time is 9h 39m. 
 Flux Conversion Factor 
 values were 516 Jy pW$^{-1}$ beam$^{-1}$ and 531 Jy pW$^{-1}$ beam$^{-1}$ at $\lambda$= 850 $\micron$ and 450 $\micron$, respectively. 
 Data were selected only when the weather conditions 
 were satisfactory for submillimeter polarimetry (0.03 \textless $\tau_{~225~ \rm{GHz}}$ \textless 0.07). Starlink software (\cite{Currieetal14}) was used for data reduction.

\subsection{ALMA polarimetry}
~~~ALMA Cycle 6 Band 3 ($\lambda$=~3.3 mm) data were utilized to investigate dust continuum polarization taken with 12-m array in January 2019  (Project code: 2018.00343.1.S). 
Data were reduced using the Common Astronomy Software Application (CASA) package (\cite{McMullinetal07}).  The synthesized beam of the data is 2.$\arcsec$958 $\times$ 2.$\arcsec$276.

 \subsection{Data selection for Spectral Energy Distribution (SED) analysis}
~~~Astropy \texttt{sedfitter} (\cite{Robitailleetal07},  2017) package was used to assess the Spectral Energy Distribution (hereafter SED) of the L1521\,F core.
For this purpose, the JCMT data described above were used along with Herschel/PACS data taken at $\lambda$= 70 $\micron$  (\cite{Bulgeretal14}), 100 $\micron$, and 160 $\micron$ (\cite{Sadavoyetal18}),  Herschel/SPIRE data of bands at $\lambda$= 250 $\micron$, 350 $\micron$, and 500 $\micron$ (\cite{Andreetal10}), IRAM 30 m MANBO-2 data taken at $\lambda$=  1.2 mm  (\cite{Tokudaetal16}), and ALMA data at $\lambda$= 3.3 mm. Fluxes at each band were measured using aperture photometry using \texttt{apphot} of the IRAF package in order to subtract background emissions.  
To derive the respective fluxes at each band, a single aperture of 6$^{\prime\prime}$ was applied. Note that the beam sizes of the data at each band vary, as shown in Table 1.  The aperture size of 6\arcsec is required for this analysis in order to set the resultant flux contribution coming from the protostar itself, not from the reflection nebula that will appear in multiple infrared bands. The flux densities of each band are summarized in Table \ref{tab:flux}. 
Bands longer than $\lambda$= 70 $\mu$m, with the exception of the $\lambda$= 3.3 mm band, were used to 
determine the overall properties of the dense core.  
Flux densities measured at $\lambda$= 3.3 mm were not used because this is the only data taken with an interferometer and it suffers significantly from a missing flux issue.  

\begin{table}
\tbl{Summary of data set for photometry}{%
  \begin{tabular}{cccc}
  \hline              
  Wavelength [$\micron$] & Instrument & Flux [mJy] & Beamsize [$^{\prime\prime}$]\\ 
  \hline
  70 & PACS & 387.65 $\pm$ 5.80 & 5.7\\
  100 & PACS & 974.30 $\pm$ 0.42 & 7.1\\ 
  160 & PACS & 1170.18 $\pm$ 11.95 & 11.2\\
  250 & SPIRE & 1307.32 $\pm$ 64.10 & 18.2\\
  350 & SPIRE & 1030.86 $\pm$ 46.10 & 25.0\\
  450 & SCUBA2 & 641.41 $\pm$ 67.10 & 7.9\\
  500 & SPIRE & 613.53 $\pm$ 45.18 & 36.1\\
  850 & SCUBA2 & 118.57 $\pm$ 3.19 & 14.1\\
  1200 & MANBO2 & 90.98 $\pm$ 2.90 & 14.0\\
  3300 & ALMA  &  0.58 $\pm$ 0.084 & 3.3 $\times$ 2.3 \\
  \hline
  \end{tabular}}\label{tab:flux}
\end{table}

\section{Results} 
\subsection{Submillimeter dust continuum polarimetry}

\begin{longtable}{*{7}{l}}
\caption{Polarization data measured with JCMT SCUBA-2/POL-2 at $\lambda$= ~850 $\micron$ and 450 $\micron$}\label{tab:segments}
\hline
  ID & Band & RA [h  m  s] & Dec [$^\circ$  '  $^{\prime\prime}$ ] & $I$ [mJy beam$^{-1}$] & $P$ [$\%$] & $ANG$ [$^\circ$] \\ 
\hline
\endfirsthead
\hline
  ID &  & RA [h  m  s] & Dec [$^\circ$  '  $^{\prime\prime}$ ] & $I$ [mJy beam$^{-1}$] & $P$ [$\%$] & $ANG$ [$^\circ$] \\  \\
\hline
\endhead
\hline
\endfoot
\hline
\multicolumn{7}{l}{ } \\

\endlastfoot
  1 & 850$\mu$m & 04:28:38.605 & +26:51:03.30 & 33.07 $\pm$ 0.93 & 14.0 $\pm$ 2.0 & -76.3 $\pm$ 4.7\\
  2 & 850$\mu$m & 04:28:38.605 & +26:51:15.00 & 59.74 $\pm$ 1.65 & 5.3 $\pm$ 1.3 & -86.3 $\pm$ 6.7\\
  3 & 850$\mu$m & 04:28:39.501 & +26:51:15.00 & 107.09 $\pm$ 1.8 & 3.2 $\pm$ 0.8 & -60.7 $\pm$ 7.2 \\
  4 & 850$\mu$m & 04:28:38.754 & +26:52:15.00 & 62.59 $\pm$ 1.48 & 5.2 $\pm$ 1.3 & -33.5 $\pm$ 8.5\\ 
  5 & 850$\mu$m & 04:28:40.548 & +26:51:51.00 & 60.08$\pm$ 2.05 & 4.8 $\pm$ 1.4 & -44.4 $\pm$ 9.0\\
  6 & 850$\mu$m & 04:28:39.651 & +26:52:03.00 & 16.78 $\pm$ 1.46 & 17.2 $\pm$ 5.0 & -81.1 $\pm$ 12.8\\
  7 & 450$\mu$m & 04:28:39.501 & +26:51:39.00 & 373.59 $\pm$ 13.01 & 7.0 $\pm$ 1.8 & 14.7 $\pm$ 10.2\\
  8 & 450$\mu$m & 04:28:39.501 & +26:51:51.00 & 132.45 $\pm$ 8.66 & 23.3 $\pm$ 5.6 & 8.1 $\pm$ 8.0\\
\end{longtable}

\begin{figure}
 \begin{center}
 \includegraphics[width=8cm]{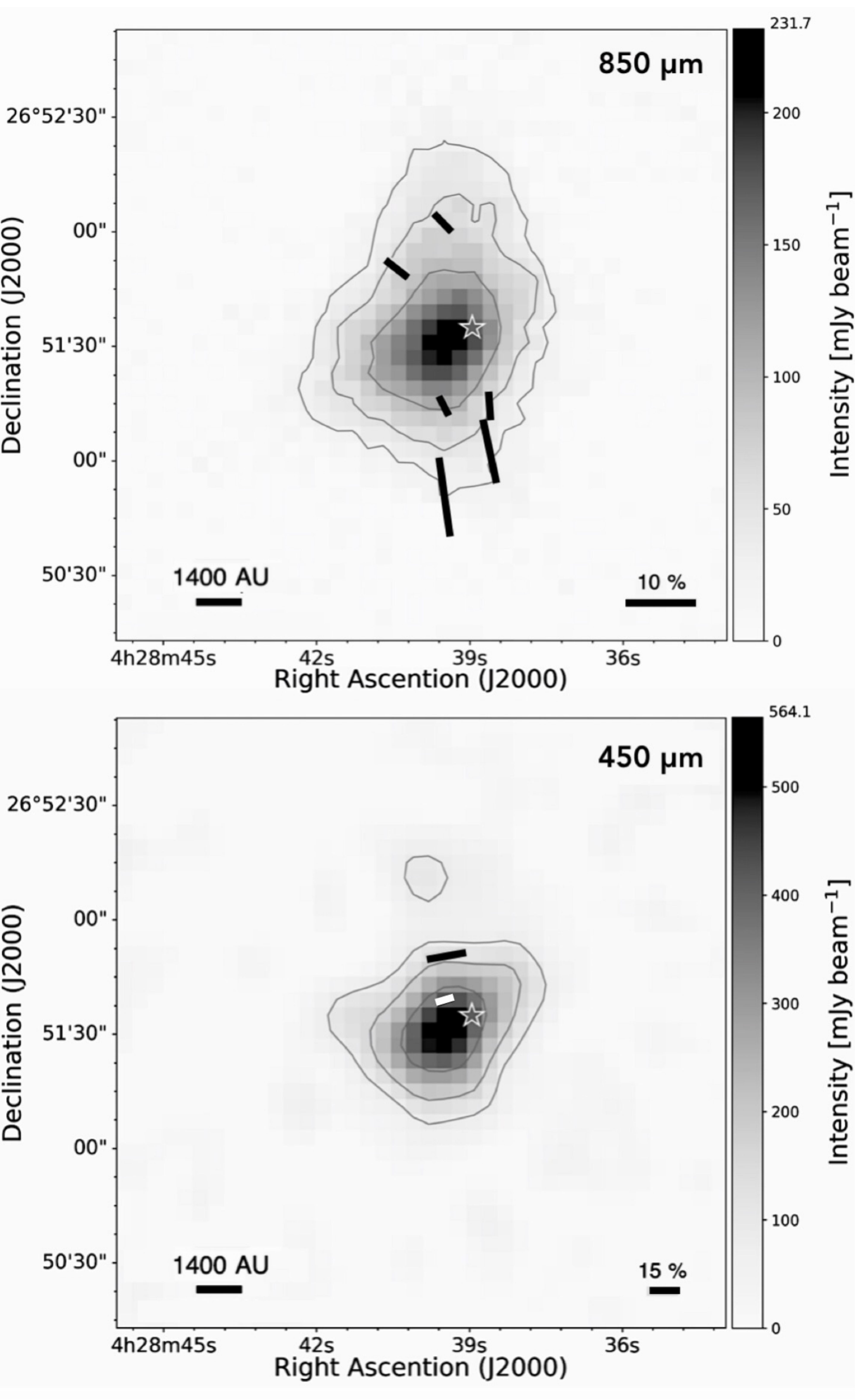}
 \end{center}
\caption{$\lambda$= 850 $\micron$ (Top panel) and 450 $\micron$ (Bottom panel) dust continuum images (one pixel corresponds to 4$^{\prime\prime}$).  
B segments (black and white thick bars) are plotted in every 12$^{\prime\prime}$ pixel$^{-1}$. 
Contours are drawn at  $5\sigma\cdot2^n$ where $n=0, 1, 2$, ~and $3$. One $\sigma$  $\sigma_{850 \micron}$ corresponds to 5 
mJy beam$^{-1}$ and $\sigma_{450 \micron}$ = 
16 mJy beam$^{-1}$.  The star on each panel marks the position of the protostar.  
}\label{fig:stImap}
\end{figure}

~~~The top and bottom panels of Figure \ref{fig:stImap} show overlays of magnetic field segments (hereafter B segments) onto the 850 $\micron$ and 450 $\micron$ continuum images, respectively. 
Black lines represent the magnetic field directions derived from polarization vectors rotated by 90$^\circ$. 
These B segments were selected according to the following criteria; signal-to-noise ratio ($S/N$) in polarized intensity ($PI$/$DPI$) \textgreater ~3 and  
$S/N$ in 
intensity  ($I$/$DI$) \textgreater ~10. 
Note that the B segments are plotted in 
12$^{\prime\prime}$  pixels in order to ensure that every data point is independent; 
this differs from the continuum images. \par 
The peak intensity of the $\lambda$= 850 $\mu$m image (Figure \ref{fig:stImap} top panel) is 232 mJy beam$^{-1}$ with rms noise levels of 1.5 mJy beam$^{-1}$.  At $\lambda$= 850 $\mu$m, the magnetic fields of the peripheral region of the core are traced and they are in an overall  
north$-$south direction, which smoothly connects to the global magnetic field direction observed with Planck at $\lambda$= 850 $\mu$m. 
The $\lambda$= 450 $\mu$m image (Figure \ref{fig:stImap} bottom panel) traces the inner warm region of the core. The corresponding peak intensity of the $\lambda$= 450 $\mu$m image is 565 mJy beam$^{-1}$ with rms noise levels of 9.9 mJy beam$^{-1}$.  
In the $\lambda$= 450 $\mu$m dust continuum component, the overall direction of the magnetic field is in the east$-$west direction, i.e., almost orthogonal to both the global magnetic field and the magnetic field directions in the peripheral region (traced at $\lambda$= 850 $\micron$) of the dense core.
Detected polarization segments at both wavelengths are summarized in Table \ref{tab:segments}.
Note that the position of the protostar, L1521\,F-IRS is at $(\alpha, \delta)_{\rm J2000.0} =$ (\timeform{4h28m38s.96}, $+$\timeform{26D51'35''}) based on ALMA $\lambda$= 0.87 mm dust continuum observations (\cite{Tokudaetal16}); this offsets from the peak flux positions at $\lambda$= 850 $\mu$m and 450 $\mu$m.

\subsection{Millimeter dust continuum polarimetry}
\begin{figure}
 \begin{center}
 \includegraphics[width=8cm]{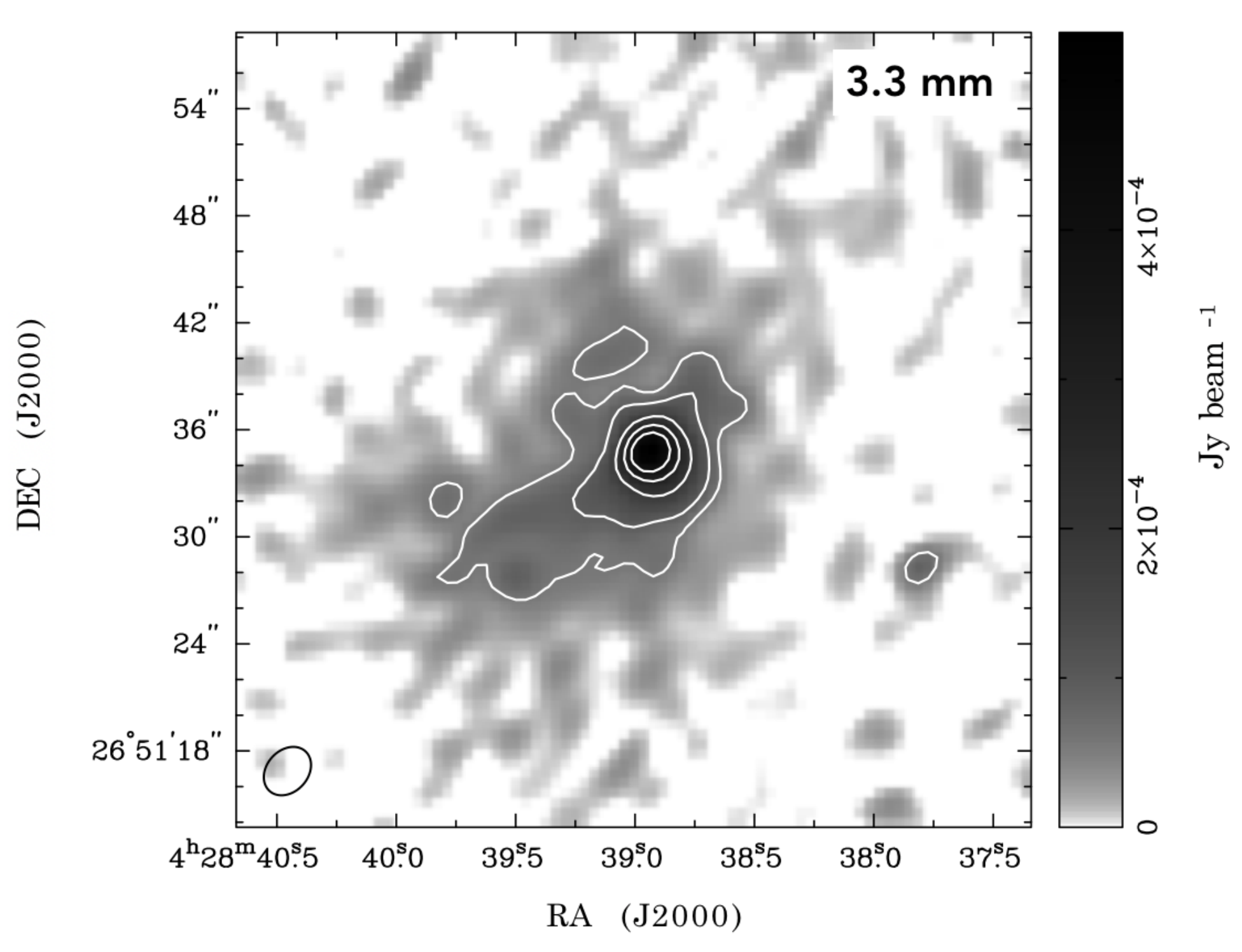}
 \end{center}
\caption{$\lambda$= 3.3 mm dust continuum image of L1521\,F taken with ALMA. White contours are drawn at levels of 3$\sigma$, 5$\sigma$, 10$\sigma$, 15$\sigma$, 20$\sigma$, where one $\sigma$ is 1.91 $\times$ 10$^{-5}$ mJy beam$^{-1}$. Spatial resolution of 
2.$\arcsec$958  $\times$ 2.$\arcsec$276   
is shown by the ellipse in the lower left corner. No significant polarization was detected in this data set. }\label{fig:alma}
 \end{figure}

~~~Figure \ref{fig:alma} shows the $\lambda$= 3.3 mm dust continuum image of L1521\,F as acquired with ALMA. 
A single continuum component is detected, with a structure elongated in the southeast direction with respect to the peak component that encases the protostar. 
Significant polarization was not detected in this dataset. 
The reason for which no significant polarization detected could either arise from a lack of sensitivity of the observation or from complex magnetic field structures in the region traced by the $\lambda$= 3.3 mm continuum. 

\section{Discussion}
\subsection{Mass estimated from dust continuum}
~~~ To estimate dense core masses traced at $\lambda$= 850 $\micron$ and 450 $\micron$, the following equation was used:  
 \begin{equation}
 M_{\rm dust}=\frac{F_\nu d^2}{{\kappa_\nu} B_{\nu}(T_{\rm dust})},
  \end{equation}
where $F_\nu$ is flux density, $d$ is distance (140 pc), $\kappa_\nu$ is dust opacity, and $B{_\nu}$($T_{\rm dust}$) is brightness (Planck function) at dust temperature $T_{\rm dust}$. 
The flux densities of $\lambda$= 850 $\micron$ and 450 $\micron$ are measured as 2.00 $\pm$ 0.13 Jy, and 9.01 $\pm$ 1.51 Jy, and 
the corresponding dust opacities 
obtained by the thin ice mantle model at these wavelengths  adopted as 1.89 cm$^2$\,g$^{-1}$ and 6.14 cm$^2$\,g$^{-1}$, respectively (\cite{Ossenkopf94}). 
Assuming typical values of $T_{\rm dust}$ = 10 K and a  gas-to-dust ratio of 100, the estimated core masses at $\lambda$= 850 $\mu$m and 450 $\mu$m are 1.36 $\pm$ 0.09 $M_{\odot}$ and 1.48 $\pm$ 0.25 $M_{\odot}$, respectively.

\subsection{SED analysis}
~~~Robitaille's model (\cite{Robitaille17}) was employed to determine the SED of the dense core in order to better assess physical properties based on unresolved observations.  As a result,  \texttt{s-ubhmi}, i.e., an SED model with a YSO was found to have the best match for L1521\,F among the large number of available models computed for 10 viewing angles at 50 apertures in a variety of evolutionary stages of star formation, as shown in Figure 3. 
The selected model considers 
following components: emissions from a central star, an Ulrich-like envelope (\cite{Ulrich76}), and a bipolar outflow cavity. The inner hole radius produced by the outflow is set as a free parameter and the dense core has an ambient medium with interstellar dust.
Table \ref{tab:sed} summarizes the physical parameters of the 
most closely matching model using this method.  For example, the estimated inclination of the bipolar outflow cavity of 88.5$^\circ$ (nearly edge-on) matches with the observation (see also Fig. \ref{fig:image}).  

 \begin{figure}
 \begin{center}
 \includegraphics[width=8cm]{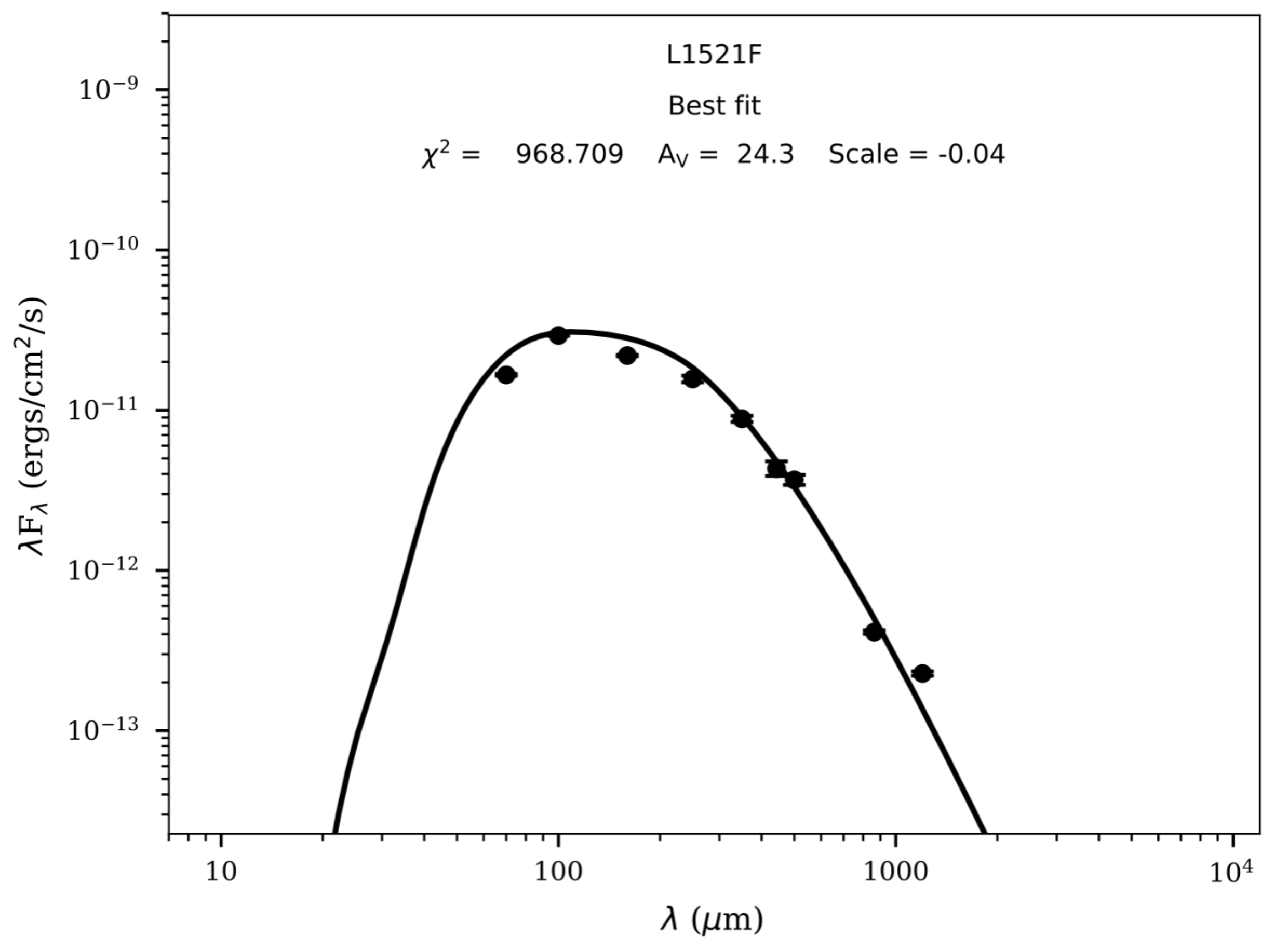}
 \end{center}
\caption{Best fit SED (black solid curve) of L1521\,F using a \texttt{sedfitter} YSO model (\cite{Robitaille17}). }\label{fig:sedfit}
 \end{figure}

 \begin{table}
  \tbl{Physical parameters estimated from SED analysis }{%
  \begin{tabular}{ccc}
  \hline              
  Parameter & unit & Result\\ 
  \hline
  star radius & $R_\odot$ & 11 \\
  stellar effective temperature & K & 2080\\
  envelope density (dust) & g/cm$^3$ & $2.78\times 10^{-20} $\\ 
  envelope centrifugal radius & AU & $4.23\times 10^{3} $ \\
  envelope inner radius & $R_\odot$ & 396\\ 
  Cavity power & - & 1.3 \\
  Cavity half-opening angle & deg & 21.4 \\
  Cavity density (dust) & g/cm$^3$ & $6.94\times 10^{-22} $ \\
  Inclination angle & deg & 88.5 \\
  of bipolar outflow cavity &     &   \\
  \hline
 \end{tabular}}\label{tab:sed}
\end{table}

\subsection{Morphology of the magnetic field within the dense core}
~~~ Soam et al. (2019) 
reported a magnetic field structure  using dust continuum polarimetry traced at $\lambda$= 850 $\micron$  with the same instrument.  
Herein, by employing an updated and improved instrumental polarization model, a better signal-to-noise (S/N) ratio with higher sensitivity was achieved in order to detect polarization, i.e., the magnetic field, which was otherwise difficult to be  detected (see section 2.1). \par

\begin{figure}
 \begin{center}
 \includegraphics[width=8cm]{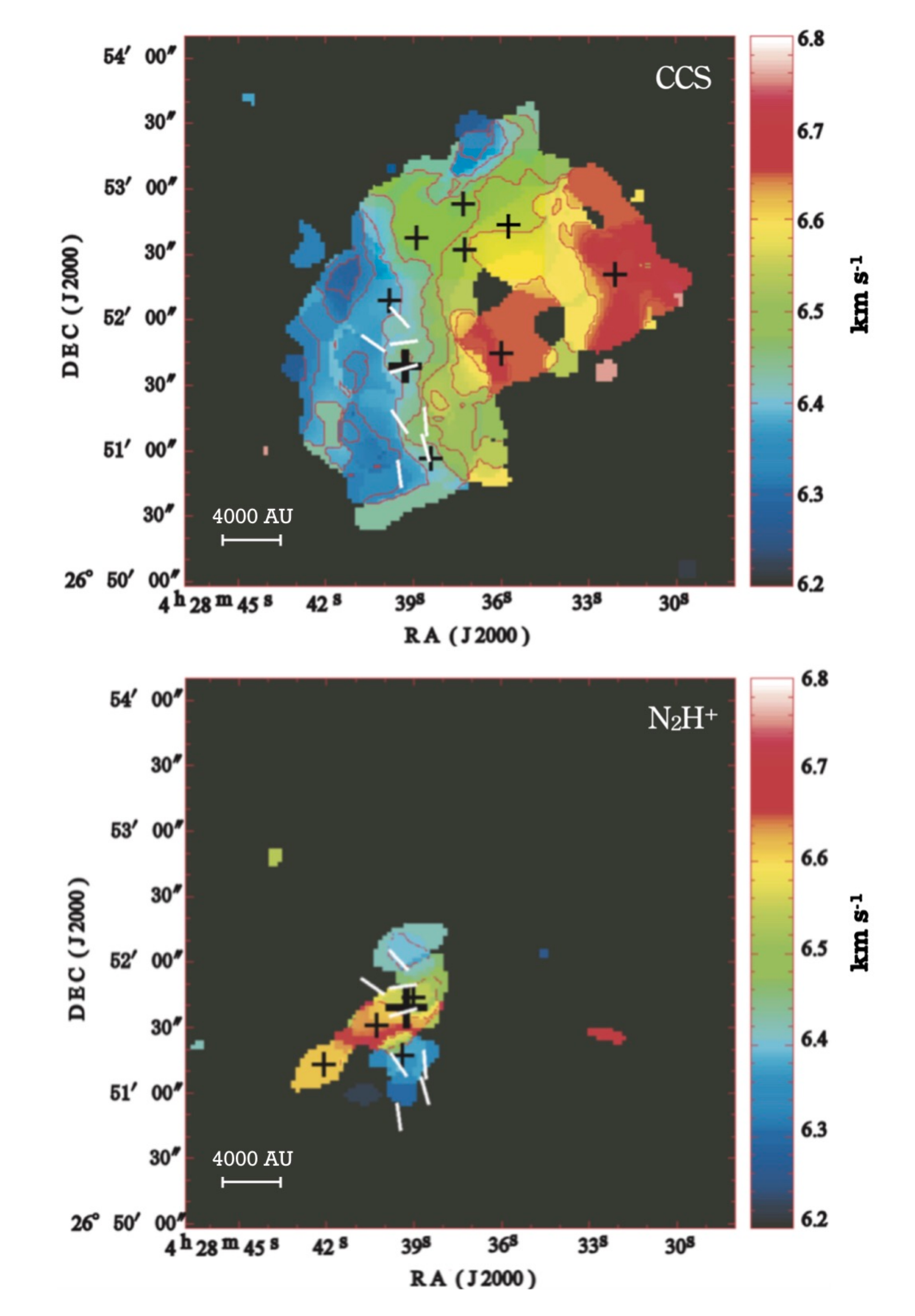}
 \end{center}
\caption{Magnetic field at $\lambda$= 850 $\micron$ and 450 $\micron$ overlaid the velocity structure from CCS (top panel) and N$_2$H$^+$(bottom panel) molecular emissions. The big cross indicates the protostar 
and small crosses indicate the positions of the substructural clumps. The length of each segment is the same.}\label{fig:ccs_n2hp}
 \end{figure}

Figure \ref{fig:ccs_n2hp} shows B segments detected at $\lambda=$ 850 $\micron$ and 450 $\micron$ overlayed onto velocity maps of CCS and N$_2$H$^+$ molecular emission reported by Shinnaga et al. (2004).  
Small cross marks in each panel show the positions of substructural clumps, while the thick cross on each panel represents the position of the protostar.  B segments detected at $\lambda$= 850 $\mu$m and 450 $\micron$ are within the clumps detected in N$_2$H$^+$. \par 
CCS molecular emissions trace the early evolutionary phase which is shown to coincide in the peripheral region of the dense core.  On the other hand, N$_2$H$^+$ molecular emissions trace the high-density, late evolutionary stage residing in the central region of the dense core.  
As such, one can confirm that the CCS peripheral region shows a velocity gradient from east to west, while the N$_2$H$^+$ central region shows a velocity gradient from northwest to southeast.  This provides a clear evidence that there is a significant gap in the velocity structure within the dense core, which has been recognized as one of the 
major outstanding research questions of such objects.  Note that magnetic field directions change drastically within and outside the central N$_2$H$^+$ clump, where a significant gap in velocity gradient is detected.

 
\subsection{Magnetic field strength and mass-to-flux ratio}
 ~~~Magnetic field strengths in L1521\,F were estimated using the Davis$-$Chandrasekhar$-$Fermi method (\cite{Davis51}; \cite{Chandrasekhar53}) measured at $\lambda$= 850 $\micron$ and 450 $\micron$. Magnetic field strength on the plane of the sky is given as
 \begin{eqnarray} 
 B_{\rm pos}= Q \sqrt{4 \pi \rho} \cdot \left( \frac{\delta V}{\delta \phi [{\rm rad}]} \right) \approx 9.3  \sqrt{n({\rm H_2})} \cdot  \left( \frac{\Delta V}{\delta\phi [^\circ]}  \right) ~\mu {\rm G}, 
 \end{eqnarray}
where $Q$ is a numerical factor introduced in the simulation by  \cite{Ostrikeretal01}, $\rho$ is the gas density in g cm$^{-3}$, $\delta V$ is the velocity dispersion in  cm\,s$^{-1}$, $\delta \phi$ is the dispersion in polarization position angles,  
$n$(H$_2$) is the number density of molecular hydrogen in cm$^{-3}$, and $\Delta V=\sqrt{8\ln2}\delta V$, which corresponds to the FWHM line width in km s$^{-1}$. Based on the results by Ostriker et al. (2001), 
we used $Q$ value of 0.5.  \par
According to the equation above, gas densities are estimated as (1.14 $\pm$ 0.08) $\times$ 10$^{-18}$ g cm$^{-3}$ and (4.5 $\pm$ 0.8) $\times$ 10$^{-18}$ g cm$^{-3}$, respectively,  
based on the $\lambda$= 850 $\micron$ and 450 $\micron$ dust continuum data. 
For velocity dispersion,  N${_2}$H$^+$ ($J = 1 - 0$) clump data of the dense core (Shinnaga et al. (2004)) 
were used, since the detected magnetic field segments are within the region where N${_2}$H$^+$ emissions were detected.   
The velocity dispersion is derived to be (0.12 $\pm$ 0.01) $\times$ $10^{5}$ cm s$^{-1}$, and the dispersion in the polarization angle is estimated by adopting the procedure described in Franco et al. (2010).  
It is found to be $20 \pm 1$ 
$^\circ$ and $11 \pm 9$ 
$^\circ$ 
measured at $\lambda$= 850 $\mu$m and 450 $\mu$m, respectively. Estimated magnetic field strengths in the 
$\lambda$= 850 $\mu$m continuum and 450 $\mu$m continuum regions were 
70 $\pm$ 10 $\mu$G and 200 $\pm$ 70 $\mu$G, respectively. Please note that the field strength derived at $\lambda$= 450 $\mu$m is based on the limited number of detected segments. 

Mass-to-flux ratios, $\mu_{\rm c}$, 
are derived by using the estimated magnetic field strengths, $B_{\rm pos}$:  
\begin{eqnarray}
 \mu_{\rm c}=\frac{(M/\Phi)_{\rm obs}}{(M/\Phi)_{\rm crit}} = 0.76\cdot\left( \frac{{N(\rm{H_2})}}{{10^{-22} {\rm cm}^{-2}}} \right) \cdot \left(\frac{{B_{\rm pos}}}{{{100 \mu {\rm G}}}} \right)^{-1}
\end{eqnarray}
 where $N$($\rm{H_2}$) is column density, $B_{\rm pos}$ is estimated magnetic field strength, and (${M/\Phi}$)$_{\rm crit}$=1/(2 $\pi$G$^{1/2}$) represents the value for magnetically critical cloud (Nakano and Nakamura 1978).
Column densities derived from the observations are (2.82 $\pm$ 0.17) $\times 10^{22}$ cm$^{-2}$ for $\lambda$= 850  $\micron$ and (7.29 $\pm$ 1.3) $\times 10^{22}$ cm$^{-2}$ for $\lambda$= 450 $\micron$.  Estimated  $\mu_{\rm c}$ values are 3.1 $\pm$ 0.2 at $\lambda$= 850$\mu$m and 2.8 $\pm$ 0.4 at $\lambda$= 450 $\micron$, respectively.  These results suggest that L1521\,F is in a 
magnetically supercritical state, i.e., gravitational forces prevail over magnetic forces.\par
Regarding the aspect that the system has multiple millimeter condensations, the mass of the protostar, MMS-1 (0.2 M$_\odot$), is dominant comparing with MMS-2 and MMS-3 (Tokuda et al. 2016, 2017). As a consequence, the magnetic flux of MMS-1 must dominate among the three condensations, considering the conservation of the mass-to-flux ratio. We expect that the influence of the less massive clumps MMS-2 and MMS-3 on the magnetic configuration is very limited. \par
 Considering their masses and distances from the central protostar, MMS-1, both MMS-2 and MMS-3 cannot significantly contribute to the twist of magnetic field lines.

\subsection{Gravitational infall scenario of L1521\,F based on comparison with resistive MHD simulations}
~~~The magnetic field structure obtained by JCMT with multi-wavelength polarimetry 
is compared with MHD simulations described by Machida et al. (2020). 
This comparison 
enables the tracing of the gravitational infall history of the puzzling dense core. Machida et al. (2020) used three-dimensional resistive magnetohydrodynamics simulations 
to address the propagation of protostellar jets and the formation of circumstellar disks as well as gravitational contraction.  
Machida et al. (2020) studied the contraction of a molecular core which has uniform magnetic ﬁeld $B_0=57 \mu$G, rigid-body rotation speed $\Omega_0$= 1.4 $\times 10^{-13} {\rm rad s^{-1}}$, and density distribution corresponding to a Bonnor-Ebert sphere (the central density $\rho_c$=2.4$\times 10^{-18} \rm{g cm^{-3}}$ and the isothermal temperature $T= 10 \rm{K}$).
They compared 9 models with different initial angles $\theta_0$ between the rotation and magnetic ﬁeld vectors.
Configurations of the system as well as magnetic field patterns are investigated for large sets of simulation results, evolution traced from the prestellar cloud phase until $\sim$560 years after the formation of a protostar. This timing corresponds to $\sim 6.3\times 10^{4}$ years since the gravitational infall initiated within the dense core. The mass of the protostar at this point is 0.04 M$_\odot$. \par
\begin{figure}
 \begin{center}
 \includegraphics[width=8cm]{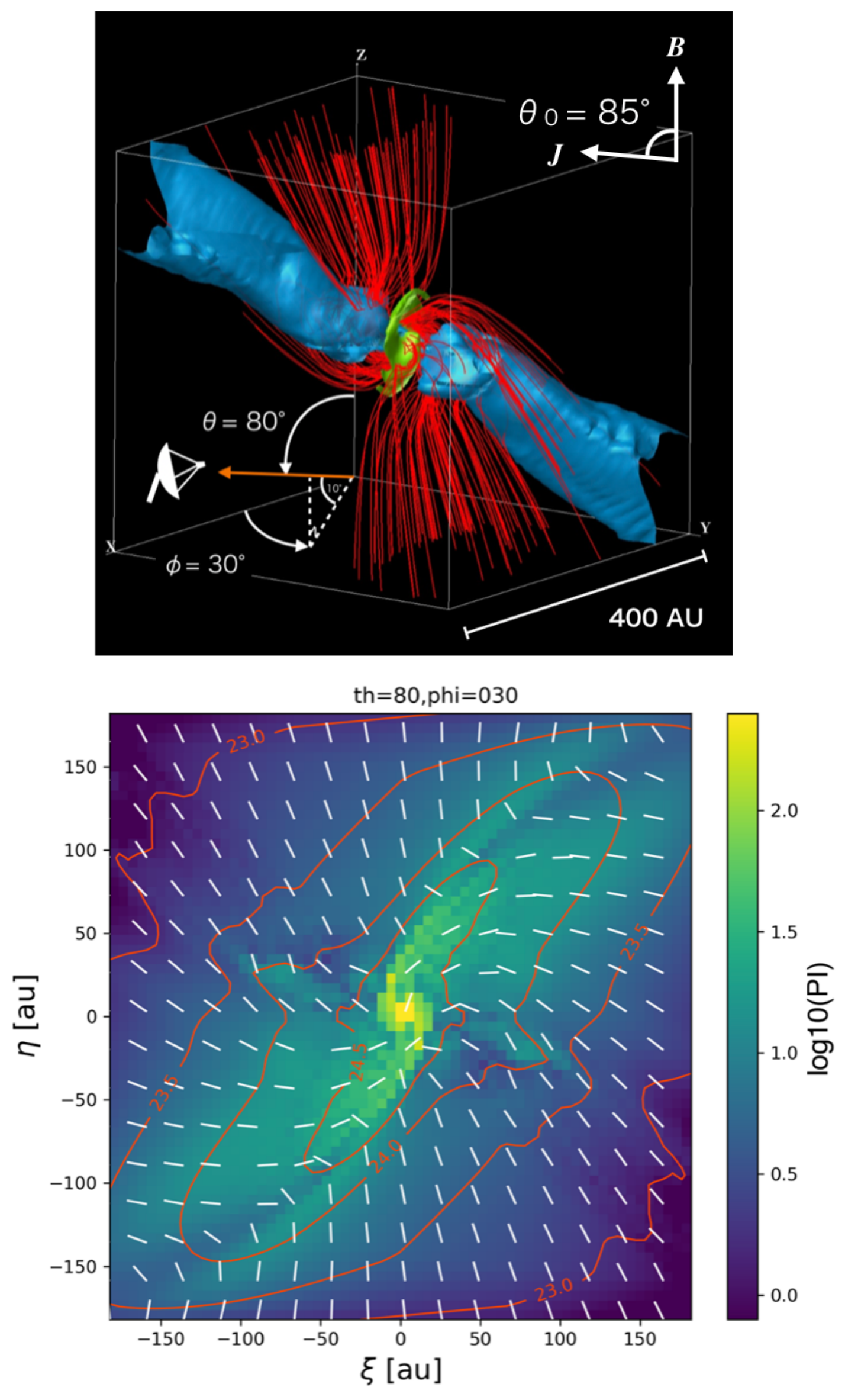}
 \end{center}
\caption{Top panel: Three-dimensional view of magnetic field lines (red lines), a pseudo-disk and high-density region (green surface) and outflow (blue surface) of the model T85 in Machida et al. 2020, in which 
initial $B$ axis and $J$ axis are misaligned by $\theta_0 = 85^{\circ}$.  
Observer's viewing 
angle is noted in orange arrow.  The initial $B$- and $J$-axes are noted at the top right corner.  Bottom panel: Expected polarized intensity (color) and surface density (contours) with B segments (white lines) viewed from 
$\theta = 80^{\circ}$ and $\phi = 30^{\circ}$. }\label{fig:mhd}
 \end{figure}

 In order to explain the north$-$south magnetic field structure detected at $\lambda$= 850 $\micron$ and the nearly orthogonal east$-$west magnetic field structure detected at $\lambda$= 450 $\micron$, we found that a model with an initial misalignment angle $\theta_0 =85^{\circ}$
 best explains the measured data.  This is the model recognized as T85 in Machida et al. (2020).  Calculating relative Stokes parameter by integrating density and magnetic field, we obtain the expected polarization pattern from the 3D MHD simulation (Tomisaka 2011; Kataoka, Machida, and Tomisaka 2012). For an observer's viewing angle, one looks at the object with a polar angle from the $z$-axis of 
 $\theta$ = 80$^{\circ}$ and the azimuth angle from $x$-axis 
 of $\phi$ = 30$^{\circ}$ (see Figure \ref{fig:mhd}). For this case, the central magnetic field is twisted by rotation of the prestellar dense cloud core and directed orthogonally to the global magnetic field, i.e.,  towards the center of the core. 
 Using this model, one can interpret that the central clump observed in N$_2$H$^+$ and the dust component detected at $\lambda$= 450 $\micron$ correspond to a pseudo-disk structure as previously shown in simulations.  L1521\,F-IRS is probably formed by processes of gravitational collapse with the rotation 
of an east$-$west angular momentum vector, which will cause the pseudo-disk to drag the magnetic ﬁeld and ﬁnally connect it to the protostar. 

\section{Conclusion}
\begin{figure}
 \begin{center}
 \includegraphics[width=8cm]{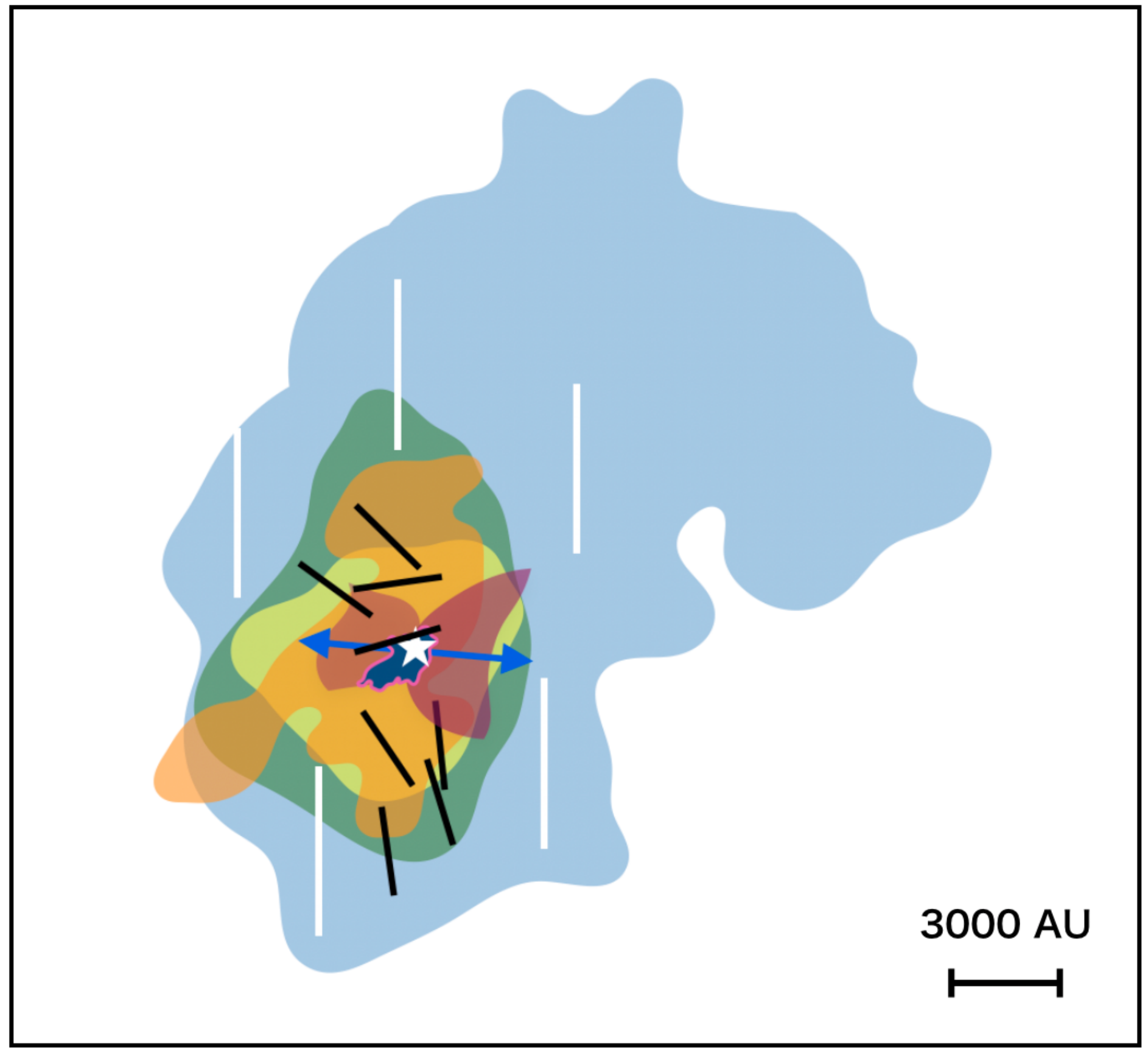}
 \end{center}
\caption{Schematic cartoon of structures detected in L1521\,F dense core. It consists of following components: sky blue : CCS; green: 850 $\micron$ continuum; yellow green: 450 $\micron$ continuum; orange: N$_2$H$^+$; red: bipolar outflow cavity; blue with pink border: 3.3 mm continuum; black lines: B segments detected at $\lambda$= 850 $\micron$ and 450 $\micron$; white lines: global magnetic field taken with Planck; white star: L1521F-IRS (the protostar); blue vectors: bipolar outflow directions emanated from the protostar. }\label{fig:image}
 \end{figure}

~~~Figure \ref{fig:image} shows a summary cartoon of L1521\,F based on the observations described in this paper.
A detailed magnetic field structure of L1521\,F is detected using JCMT, SCUBA-2 and POL-2. The magnetic field in a  peripheral cold region of the dense core threads a north$-$south direction as observed at $\lambda$= 850 $\micron$, 
while an orthogonal east$-$west magnetic field is observed at $\lambda$= 450 $\micron$ tracing the central warm region of the core.
ALMA Band 3 observation did not detect a magnetic field structure at the scale of 1500 AU. 

Estimated magnetic field strengths thus are 70 $\mu$G and 200 $\mu$G in the peripheral region traced at $\lambda$= 850 $\micron$ and in the central warm region traced at $\lambda$= 450 $\micron$, respectively. The resulting mass-to-flux ratio is found to be approximately 3 times greater than that of magnetically critical value for both regions, indicating that gravitational forces predominate over magnetic turbulence forces in the dense core. \par

Comparing observational results with MHD simulations, a  gravitational infall scenario is proposed. 
This indicates that the rotation axis of L1521\,F-IRS is misaligned with respect to initial global magnetic field axis by 85$^{\circ}$, which is almost perpendicular. One can thus interpret that the central magnetic field has been twisted by the rotation of the prestellar cloud and directed orthogonal to the global magnetic field.  The central clump traced with N$_2$H$^+$ emissions and using the $\lambda$= 450 $\micron$ dust continuum most likely corresponds to a pseudo-disk as shown in the simulation.

\begin{ack}
~~~The authors thank Drs. Tokuda and Saigo for the fruitful discussion. The James Clerk Maxwell Telescope is operated by the East Asian Observatory on behalf of The National Astronomical Observatory of Japan; Academia Sinica Institute of Astronomy and Astrophysics; the Korea Astronomy and Space Science Institute; the National Astronomical Research Institute of Thailand; Center for Astronomical Mega-Science (as well as the National Key R\&D Program of China with No. 2017YFA0402700). 
Additional funding support is provided by the Science and Technology Facilities Council of the United Kingdom and participating universities and organizations in the United Kingdom and Canada.Additional funds for the construction of SCUBA-2 were provided by the Canada Foundation for Innovation. This research also used the facilities of the Canadian Astronomy Data Centre operated by the National Research Council of Canada with the support of the Canadian Space Agency. 
This paper makes use of the following ALMA data: ADS/JAO.ALMA$\#$2018.1.00343.S. ALMA is a partnership of ESO, NSF (USA) and NINS, together with NRC, MOST and ASIAA, and KASI, in cooperation with the Republic of Chile. The Joint ALMA Observatory is operated by ESO, AUI/NRAO, and NAOJ. This work is supported in part by a Grant-in-Aid for Scientific Research of Japan (17K05388; 19H01938; 19K03919) and MEXT program supporting female and young researchers in research projects at Kagoshima University. 
This research made use of Astropy (Astropy Collaboration 2013, 2018).  Data analysis was in part carried out on the Multi-wavelength Data Analysis System operated by the Astronomy Data Center (ADC), National Astronomical Observatory of Japan. 
 %
\end{ack}



\end{document}